\documentclass[a4paper,12pt]{article}
\usepackage[top=3cm, bottom=3cm, right=2.5cm, left=2.5cm]{geometry}
\usepackage{amsmath,amsfonts,amssymb,amsthm,array,bm,bbm}
\usepackage[utf8]{inputenc}
\usepackage[english]{babel}
\usepackage{natbib}
\usepackage{caption, subcaption}
\title{The Performance of Largest Caliper Matching: A Monte Carlo Simulation Approach}
\author{Sharif Mahmood \\ {\tt sharif1@ksu.edu}\\ Department of Statistics and Actuarial Science \\ The University of Iowa}
\date{\today}
\usepackage[dvipsnames]{color}
\definecolor{maroon}{rgb}{0.5, 0.0, 0.0}
\usepackage{graphicx}
\usepackage{hyperref}
\usepackage{verbatim, animate, fancyvrb, bbding}
\usepackage{listings}
\usepackage{multirow}
\usepackage{setspace}

\theoremstyle{plain}

\begin{document}
\maketitle
\begin{abstract}
\noindent
The paper presents an investigation of estimating treatment effect using different matching methods. The study proposed a new method which is computationally efficient and convenient in implication---\emph{largest caliper} matching and compared the performance with other five popular matching methods by simulation. The bias, empirical standard deviation and the mean square error of the estimates in the simulation are checked under different treatment prevalence and different distributions of covariates. A Monte Carlo simulation study and a real data example are employed to measure the performance of these methods. It is shown that matched samples improve estimation of the population treatment effect in a wide range of settings. It reduces the bias if the data contains the selection on observables and treatment imbalances. Also, findings about the relative performance of the different matching methods are provided to help practitioners determine which method should be used under certain situations. 
\end{abstract}

\section{Introduction} \label{Introduction}
Matching methods are popular to estimate the unbiased estimate of the treatment effect both in randomized and non-randomized experiments. In randomized experiment, researchers use matching methods to form pair/block similar subjects and assign treatments. In non-randomized experiment, researchers use pretreatment covariates to match the treated subjects with control subjects and attempt to replicate a randomized experiment as if the treatments were randomly assigned. When the covariate distributions of the treated and control subjects are different---crude analysis could make a substantial bias. An appropriate matching method should reduce bias due to covariates by reducing the observed and unobserved covariate imbalances between treated and control groups.  

There are plenty of matching methods that have been developed in literature that improve the covariate balance iteratively by estimating a distance between treated units and potential controls, finding the matches, and checking balance until a satisfactory level is achieved. 
When there are large number of covariates---it may not possible to reduce the imbalance of all covariates altogether.  
The goal can be achieved by propensity score matching of treated and control groups that reduce bias due to the covariates \citep{Rosenbaum1983Central, Dehejia1999Causal}. In contrast, propensity score matching has been challenged as a matching method that can increase imbalance if the propensity score model is misspecified \citep{Diamond201Genetic, KingWhy}. 
Another common approach that can reduce the imbalance between treated and control groups is Euclidean/Mahalanobis distance matching. One limitation of such distance metric is that if there is an extreme outlier in one covariate for a unit---the estimated variance for that covariate will be high, and Euclidean/Mahalanobis distance ignore the differences in that covariate. In a special case, \citet{Gu1993comparison} reported that if a binary covariate  that takes values 1 and 0 with probabilities $p$ and $1-p$; whenever $p\to 0$, Mahalanobis distance tries to match a rare treated unit with this covariate equal to 1. 
The performance of Mahalanobis distance matching in better in small data whereas the performance of propensity score matching better in large data set. Once the matched sample is selected through distance metric, very simple methods can be used to analyze the outcomes, and typical analysis of matched samples do not require the parametric assumptions of most regression methods \citep{Rosenbaum1985Cons}. 

The quantity of interest for the outcome analysis depends on the researcher objectives---for continuous response the most common estimand is average treatment effect (ATE) or average treatment effect on the treated (ATT) and odds ratio for the binary outcomes. Note that if a matching method that discards both treated and control units to find a fine balance---do not result ATE or ATT.  In this article, we focus on ATT to compare the performance of the estimation of largest caliper matching compare with other matching methods. 

Section \ref{description} describes matching methods that have been considered in this article. Section \ref{montecarlo} describes a series of Monte Carlo simulations to examine the performance of these methods in estimating treatment effects. Particularly, we report on bias, standard deviation and mean square error (MSE) of the estimates. 
Section \ref{casestudy} presents analysis of the right heart catheterization data. Finally, in Section \ref{conclusion}, we summarize our findings.

\section{Methods} \label{description}

Several researches have been conducted to compare the matching methods. \citet{Elze345} compared four propensity score matching methods to covariate adjustment on four cardiovascular observational studies. \citet{Austin2014comparison} compared 12 matching methods for 1:1 matching on the propensity score. \citet{Kewei2000substantial} observed that substantially greater bias reduction is possible if the number of controls in match to each treated unit is not fixed. \citet{Gu1993comparison} compared optimal matching with nearest neighbor matching based on Mahalanobis distance. In this article, we considered six different matching methods: nearest neighbor matching with replacement (NNWR), nearest neighbor matching without replacement (NNWOR), optimal matching (OPT), full matching (FL), genetic matching (GM) and largest caliper matching (LC). The choice of selecting a matched sample differs in the methods and each serves to achieve specific objectives. 

\subsection{Nearest Neighbor Matching With Replacement}
NNWR matching matches all treated subjects to their nearest control subjects based on a distance metric. In this method, each treatment subject can be matched to the closest control subject, even if that control subject is matched more than once. Because this approach can provide closer matches on the distance than nearest-available matching without replacement, it can be beneficial for reducing bias in the analysis. In our analysis, we used Mahalanobis distance metric to find the nearest control for the treated subjects. An illustration of the method is shown in Figure \ref{fig:nnwor}.

\subsection{Nearest Neighbor Matching Without Replacement}
NNWOR requires that each match contains exactly one treated subject and exactly one control subject also known as 1:1 match. Once a control subject is matched with a nearest treated subject that control subject is no longer eligible for consideration as a match for other treated subjects. 
That is why, NNWOR is also known as ``greedy" matching. It can be beneficial when there are enough good matches. Mahalanobis distance metric is used in the analysis to find nearest control for the treated subjects. 
The method is illustrated in Figure \ref{fig:nnwr}. 

\subsection{Optimal Matching without Replacement}
The optimal matching method seek to match subjects to minimize a global discrepancy measure, like the sum of distances within matched sets \citep{Rosenbaum1989Optimal}. \citet{Greevy2004Optimal} develops the idea to improve matching methods with the goal of optimizing the overall similarity of matched subjects. It does not make any difference if optimal matching with replacement compared to nearest neighbor with replacement. In our analysis, exactly one treated subject is matched with exactly one control subject that minimizes overall Mahalanobis distance. Figure \ref{fig:optmatch} illustrates the method.

\subsection{Full Matching}
Full matching considers that there exist at least one matched control (treated) subject for every treated (control) subject. Again, the treated (control) subjects are not matched with the matched control (treated) subjects. One can choose $1:k$ or $k:1$ matching in full matching. The flexibility of this matching method can result in using more of the data at hand and yield more effective comparisons (in terms of effective sample size) and closest-possible matches on any given distance \citep{Hansen2004FullMatching}. In our analysis, we considered 1:3 matching in full matching with calipers of width equal to 0.2 of the standard deviation of the logit of the propensity score. The choice of the ratio was based initial performance before we conduct the whole simulation. Figure \ref{fig:full} illustrates how the subjects would be matched using the method.

\subsection{Genetic Matching}
\begin{figure}[!ht]
\begin{center}
    \begin{subfigure}[b]{0.31\textwidth}
        \includegraphics[width=\textwidth]{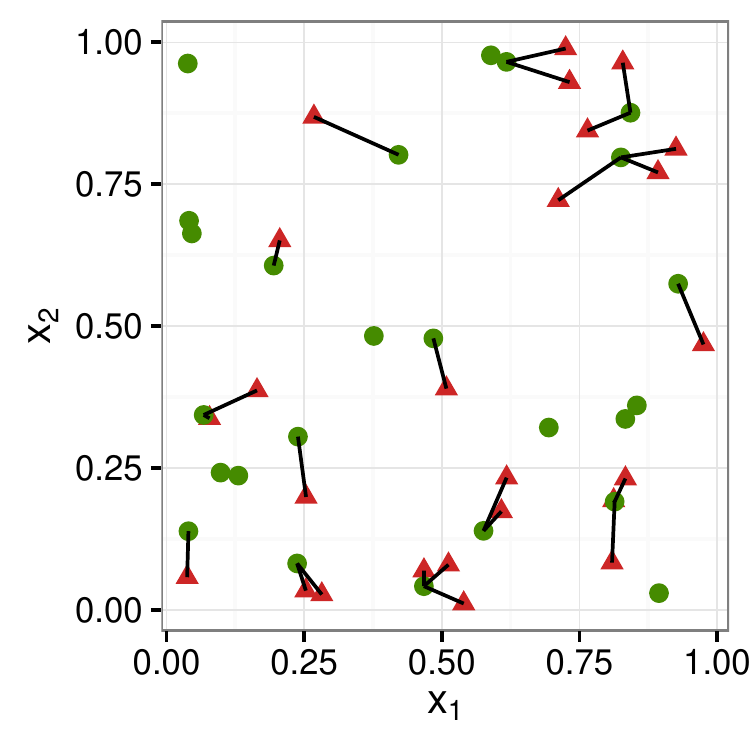}
        \caption{}%
        \label{fig:nnwor}
    \end{subfigure}
    \begin{subfigure}[b]{0.31\textwidth}
        \includegraphics[width=\textwidth]{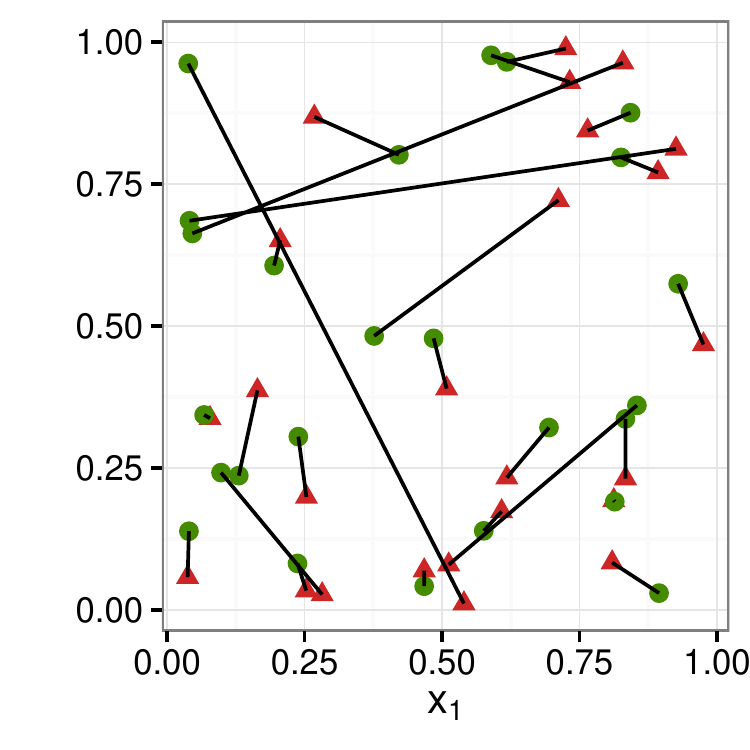}
        \caption{}
        \label{fig:nnwr}
    \end{subfigure}
    \begin{subfigure}[b]{0.31\textwidth}
        \includegraphics[width=\textwidth]{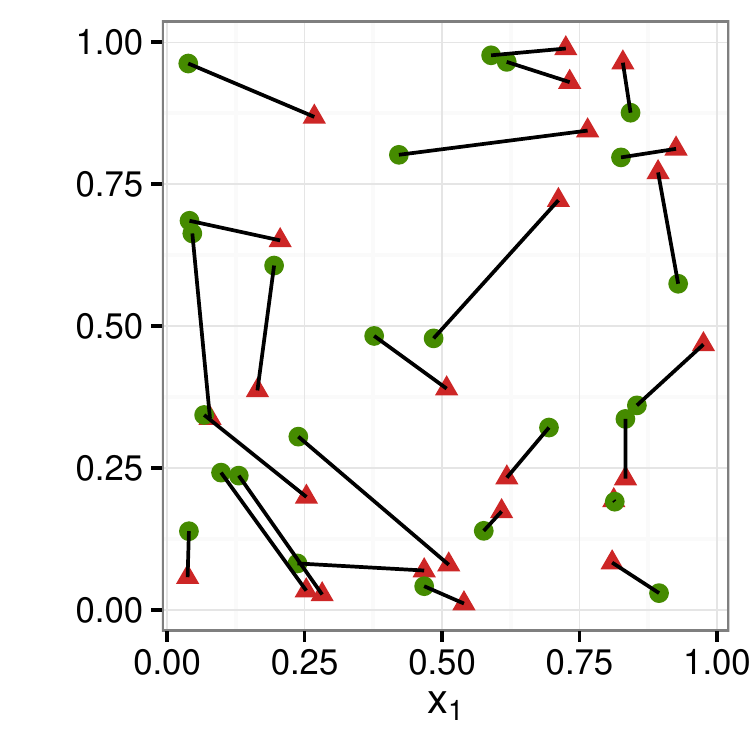}
        \caption{}
        \label{fig:optmatch}
    \end{subfigure}

    \begin{subfigure}[b]{0.31\textwidth}
        \includegraphics[width=\textwidth]{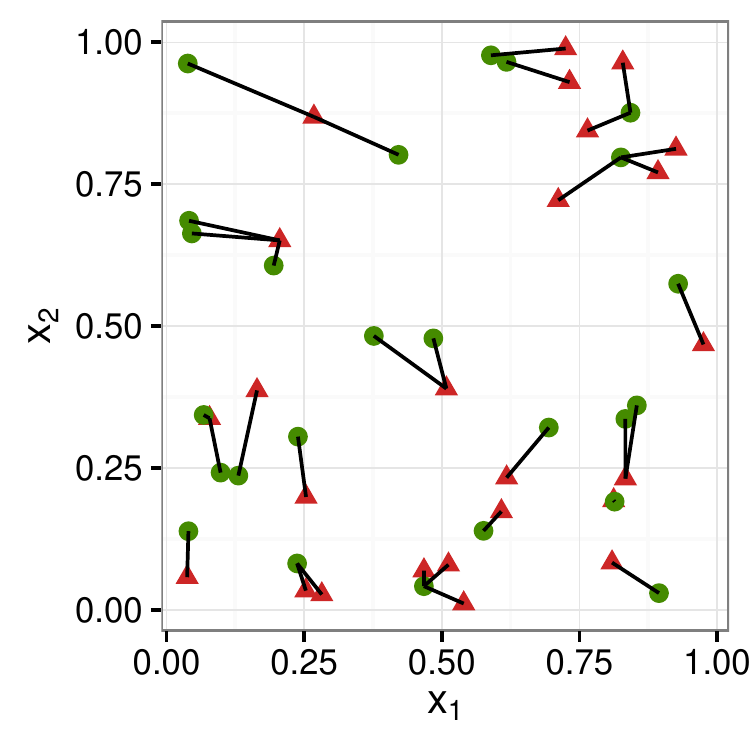}
        \caption{}
        \label{fig:full}
    \end{subfigure}
    \begin{subfigure}[b]{0.31\textwidth}
        \includegraphics[width=\textwidth]{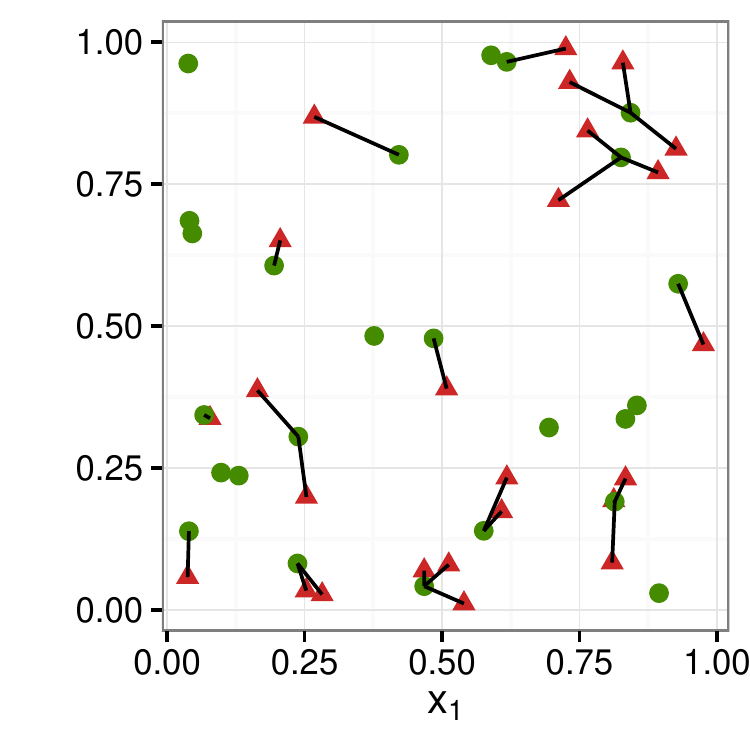}
        \caption{}
        \label{fig:genmatch}
    \end{subfigure} 
    \begin{subfigure}[b]{0.31\textwidth}
        \includegraphics[width=\textwidth]{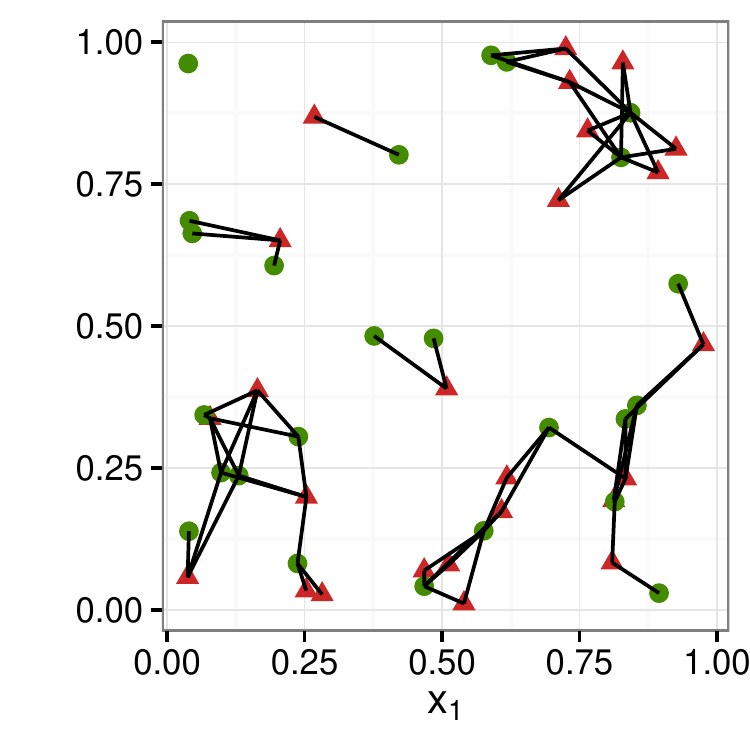}
        \caption{}
        \label{fig:lc}
    \end{subfigure} 
\end{center}    
\caption{Illustration of different matching methods. The sample consists of 50 subjects, both treated and control groups have 25 subjects each. We observe two covariates $x_1$ and $x_2$, for each subject. The red triangles indicate treated subjects and green circles indicate control subjects. Edges (based on Mahalanobis distance) indicate matched groups. A good matching method should avoid long edges, as they corresponds to increase covariate imbalance.}
\label{fig:match}
\end{figure}

\citet{Diamond201Genetic} proposed genetic matching that automates the iterative process of checking and improving overall covariate balance to determine the given covariates' weight and ensures convergence to the optimal matched sample. They proposed a distance metric for the method that minimize the overall imbalance by minimizing the largest individual discrepancy based on $p$-values from paired $t$-test. Figure \ref{fig:genmatch} presents the sample that would be matched using the method.

\subsection{Largest Caliper Matching} \label{lcc_match}
We introduce a method that provides a heuristic approach to select the maximum amount of imbalance that researchers want to accept for a match given a covariate, namely largest caliper matching. 
For largest caliper matching we consider the following distance metric:
\begin{equation}
D^*({x}_{ip}, {x}_{i^\prime p}) = \max_{p} \frac{|x_{ip} - x_{i^\prime p}|}{c_p}. \label{eq:mydist}
\end{equation}
$D^*({x}_{ip}, {x}_{i^\prime p})$ is the amount of dissimilarity between $i$th treated and $i^\prime$th control subject. Here $c_p$ is a research-selected parameter for how much imbalance on covariate $p$ is acceptable for a match. For example, if researchers want to match a treated subject of age 40 with a control subject of age within 35 and 45, then in this case $c_p = 5$. Similarly, if researchers want to match with a male treated subject with a female control subject then $c_p=1$. A large value of $c_p$ ensures a large number of matched subjects. For $k$ categories, one can make $k-1$ dummy variable and match in terms of the reference category or give weights to the units based on the proportions of categories. If $D^*({x}_{ip}, {x}_{i^\prime p}) \leq 1$, then we say that an acceptable match. All the matched units that have at least one acceptable match as described in Figure \ref{fig:lc}---form a cluster of homogeneous subjects---are analyzed giving weights to the clusters based on the subjects in that cluster by total subjects. 

We note several importance of largest caliper matching: First, largest caliper matching match based on the amount of imbalance that researchers want to accept for a covariate. For example, $D^*({x}_{ip}, {x}_{i^\prime p})=0$ means exact match based on $p$th covariate that researchers want to use for matching. Again, $D^*({x}_{ip}, {x}_{i^\prime p})=\infty$ means match on $p$th covariate is negligible. Often it is not possible to reduce the imbalance for every covariate altogether, equation \eqref{eq:mydist} might not be optimal by random choice of the $c_p$. We recommend choosing the $c_p$ based on the important covariates that are related to the treatment assignment and study outcome. For large data set one can consider $c_p$ as the caliper for the propensity score \citep{Lunt2014}. \citet{Austin2011Optimal} observed optimal calipers of width equal to 0.2 of the standard deviation of the logit of the propensity score when estimating differences in means and differences in proportions in observational studies. Second, largest caliper matching is a heuristic matching method---for a given $c_p$---the average run time of the method is faster than optimal matching. Third, the choice of $c_p$ could be based on the quantity of interest. For example, if the quantity of interest is average treatment effect for the treated (ATT) (average treatment effect for the control (ATC)), then we chose the $c_p$ in such a way so that every treated (control) subject has at least one matched control (treated) subject. Fourth, largest caliper matching is a version of cardinality matching, where within a given balance of the covariates, the maximum number of units that can be considered for analysis are considered. Finally, largest caliper matching ensures to discard the extreme subjects that can increase the substantial bias in the analysis \citep{King2006danger}.

\section{Monte Carlo Simulations} \label{montecarlo}
The simulations performed in the current paper are simplistic matching simulations proposed in the literature  \citep{Austin2014comparison, Pirracchio2015}.
We conduct a number of Monte Carlo simulations to compare the performance of six matching methods on binary outcome. In each simulated sample, we compute an estimate $\hat{\tau}$ of the true parameter $\tau$. We assessed the performance of each method using the following three criteria: 
\begin{itemize}
\item Bias in estimating treatment effects: $\bar{\tau} -\tau$ where $\bar{\tau}=\sum_{l=1}^N \hat{\tau}/N $ and $N$ is the total number of simulation.
\item Standard deviation of the estimated treatment effect: $\sqrt{\sum_{l=1}^N (\hat{\tau} - \bar{\tau})^2/(N-1)}$.
\item Mean square error of estimated treatment effects: $\sqrt{\sum_{l=1}^N (\hat{\tau}-\tau)^2/N}.$ 
\end{itemize}

\subsection{The Setup}
We considered $\bm{X}$ be a vector of 5 covariates that had effect both on the treatment assignment and the outcome. The treatment assignment model was generated from a linear combination of the covariates:
$$\text{logit}(\pi_{t})=\beta_{0,t} + \beta_1 x_{1} + \beta_2 x_{2} + \beta_3 x_{3} +\beta_4 x_{4} +\beta_5 x_{5},$$  
where $ \bm{\beta} = (\beta_1, \beta_2, \beta_3, \beta_4, \beta_5)  = (\log(1.25),  \log(1.5), \log(1.75), \log(2),\log(2))$. Thus, there were one covariate that had a weak effect on each of treatment effect and outcomes, one covariate had a moderate effect on each treatment assignment and outcomes, one covariate that had a strong effect on each of treatment assignment and outcomes, and two covariates that had a very strong effect on both treatment assignment and outcomes.  
The intercept of the treatment assignment model ($\beta_{0,t}$) was generated so that the proportion of subjects in the simulated sample that were treated was fixed at a desired proportion. 
We assigned treatment status (denoted by $z$) of subjects from a Bernoulli distribution with parameter $\pi_{t}$. 
The dichotomous outcome was generated using the following logistic model: 
$$\text{logit}(\pi_{\text{o}}) = \beta_{0,\text{o}} + \tau z+ \beta_1 x_{1} + \beta_2 x_{2} + \beta_3 x_{3} +\beta_4 x_{4} +\beta_5 x_{5}.$$ 
We then generated a binary outcome for each subject from a Bernoulli distribution with parameter $\pi_{\text{o}}$. We selected the intercept, $\beta_{0,\text{o}}$, in the logistic outcome model so that the incidence of the outcome would be approximately 0.10 if all subjects in the population were control. In a given simulated data set, we simulated a binary outcome for each subject, under the assumption that all subjects were not treated ($z=0$). We then calculated the incidence of the outcome in the simulated data set. 

We selected the conditional log odds ratio $\tau$ so that average odds in treated subjects' due to treatment would be approximately 0.5. 
The same value of $\tau$ was used to generate a cohort of $n=5000$ in a given scenario. Because we were simulating data with a desired ATT, the value of $\tau$ would depend on the proportion of subjects that were treated. This approach allows for variation in subject-specific treatment effects. The logistic model is used to simulate data with an underlying average treatment effect in the treated because such an approach will guarantee that individual probabilities of the occurrence of the outcome will lie within [0,1].

In Monte Carlo simulation, we consider a complete factorial design in which the following two factors were allowed to vary: (1) the distribution of the 5 pretreatment covariates; (2) the proportion of subjects that received the treatment. 
We considered four different distributions for the 5 pretreatment covariates: 
(i) the 5 covariates had independent standard normal distributions; 
(ii) the 5 covariates were from a multivariate normal distribution. Each variable had mean zero and unit variance, and the pair-wise correlation between variables was 0.25; 
(iii) the first two variables were independent Bernoulli random variables each with parameter 0.5, whereas the other three variables were independent standard normal random variables; 
(iv) the 5 random variables were independent Bernoulli random variables, each with parameter 0.5. 
For the second factor, we considered five different levels for the proportion of subjects that were treated: 0.1, 0.15, 0.2, 0.25, 0.3 and 0.35. Hence, there are 24 different scenarios of the study:  four different distributions for the pretreatment covariates times six levels of the proportion of subjects that were treated.

In each of the 24 scenarios, we simulated $N=1000$ datasets, each consisting of $n=5000$ subjects. There were two reasons to use simulated datasets of size 5000. First, matching methods can be computationally intensive for large data. We considered a moderate size of the data that are available in real life, e.g. SUPPORT data. Second, researchers in different field usually have different size of the data---we observed in most cases these methods have been used in datasets of size around 5000. From the setup, we know the important covariates (i.e. $x_4$ and $x_5$) when matching and a good matching method should have more weight on these covariates.
Though in real life it is unknown that which variables are important for treatment and outcome but in practice---researchers use the existing literature or subject-matter knowledge and expertise to identify important variables that affect the treatment assignment or outcome. 
In each matched sample, we estimated the log odds ratio as the treatment effect. As the matched sample removes the effect of confounding due to pretreatment covariates---it was expected the estimates were unbiased. 

In SUPPORT data, we check the covariate imbalance by standardized difference. For continuous variables, the standardized difference is defined as $d=(\bar{x}_t - \bar{x}_c)/ \sqrt{(s^2_t + s^2_c)/2}$, 
where $\bar{x}_t$ and $\bar{x}_c$ denote the sample mean of the covariate in treated and control subjects, respectively, whereas $s^2_t$ and $s^2_c$ denote the sample variance of the covariate in treated and control subjects, respectively. For dichotomous variables, the standardized differences are defined as $d=(\hat{p}_t - \hat{p}_c)/\sqrt{(\hat{p}_t (1-\hat{p}_t) + \hat{p}_c (1-\hat{p}_tc))/2},$ where $\hat{p}_t$ and $\hat{p}_c$ denote the prevalence or mean of the dichotomous variable in treated and control subjects, respectively. 


\subsection{Results} \label{simulation}
\begin{figure}[!ht]
\begin{center}
    \begin{subfigure}[b]{0.25\textwidth}
        \includegraphics[height = 3.5cm, width = 3.5cm]{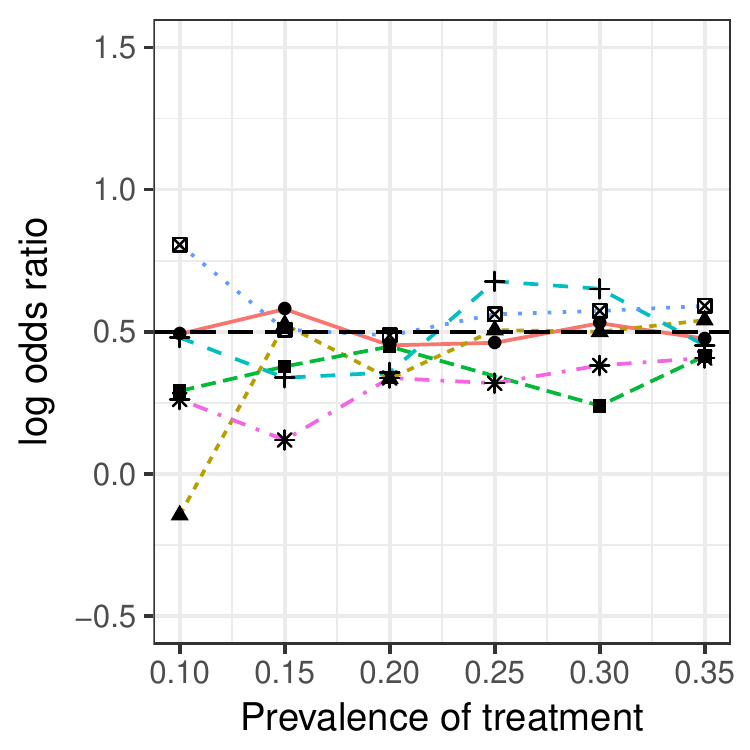}
        \caption{}
        \label{fig:odds_norm}
    \end{subfigure}
    ~ 
    \begin{subfigure}[b]{0.25\textwidth}
        \includegraphics[height = 3.5cm, width = 3.5cm]{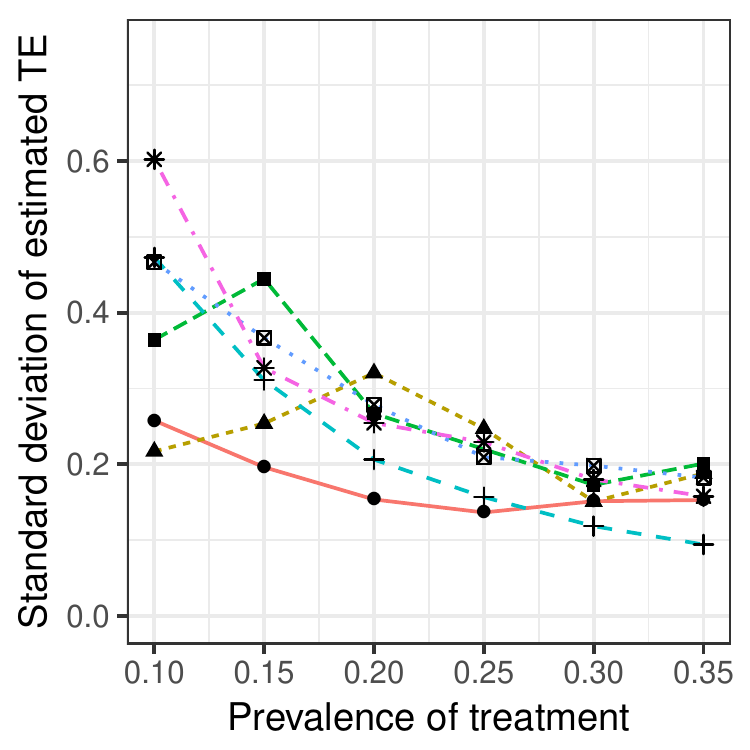}
        \caption{}
        \label{fig:std_norm}
    \end{subfigure}
    \begin{subfigure}[b]{0.25\textwidth}
        \includegraphics[height = 3.5cm, width = 3.5cm]{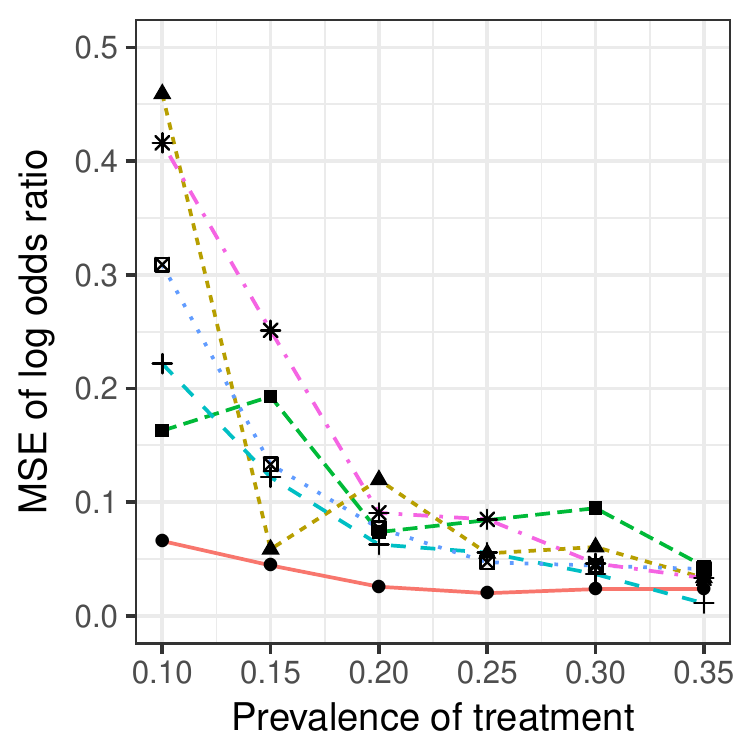}
        \caption{}
        \label{fig:mse_norm}
    \end{subfigure}
        \begin{subfigure}[b]{0.1\textwidth}
        \includegraphics[height = 2.7cm, width = 1.2cm]{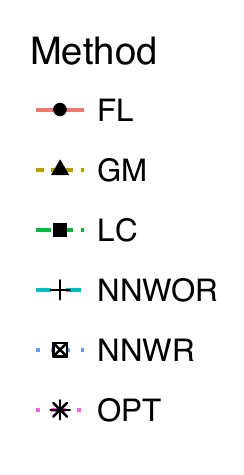} \vspace{1cm}
    \end{subfigure}    
\end{center}    
\caption{Treatment effect: log odds ratio, standard deviation of estimated log odds ratio and mean squared error of log odds ratio under independent normally  distributed covariates.} \label{fig:norm}
\end{figure}

In Figure \ref{fig:norm} we report the log odds ratio, standard deviation and mean square error of the log odds ratio when the pretreatment covariates were independently normally distributed. Figure \ref{fig:odds_norm} shows the bias of the methods under different treatment prevalence. A horizontal line has been added to each panel denoting the magnitude of the true log odds ratio 0.5. Figure \ref{fig:std_norm} and \ref{fig:mse_norm} show the standard deviation and mean square error of the estimated log odds ratio, respectively. In general, as the prevalence of treatment increased the precision of the estimates increased for all matching methods. Optimal matching and nearest neighbor matching with/without replacement tended to have similar performance under independently normally distributed covariates. Amongst all methods, 1:3 full matching with caliper showed less standard deviation and mean square error of the estimated log odds. Largest caliper matching was the second choice in this scenario. Note that when the treatment prevalence was small, e.g. 10\%, 1:1 nearest neighbor with/without replacement or optimal matching discarded at least 80\% of the subjects from the data. 

\begin{figure}[!ht]
\begin{center}
    \begin{subfigure}[b]{0.25\textwidth}
        \includegraphics[height = 3.5cm, width = 3.5cm]{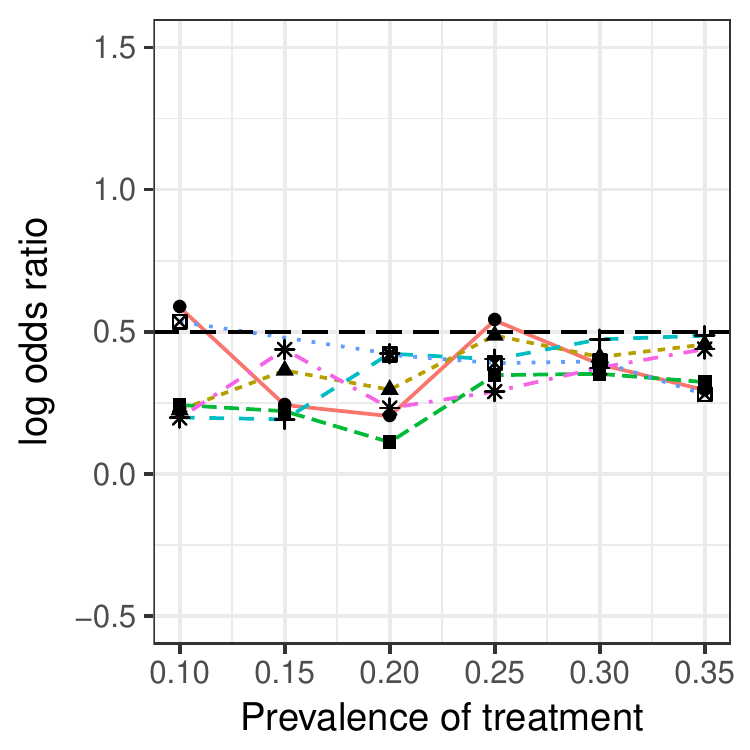}
        \caption{}
        \label{fig:odds_multi}
    \end{subfigure}
    ~ 
    \begin{subfigure}[b]{0.25\textwidth}
        \includegraphics[height = 3.5cm, width = 3.5cm]{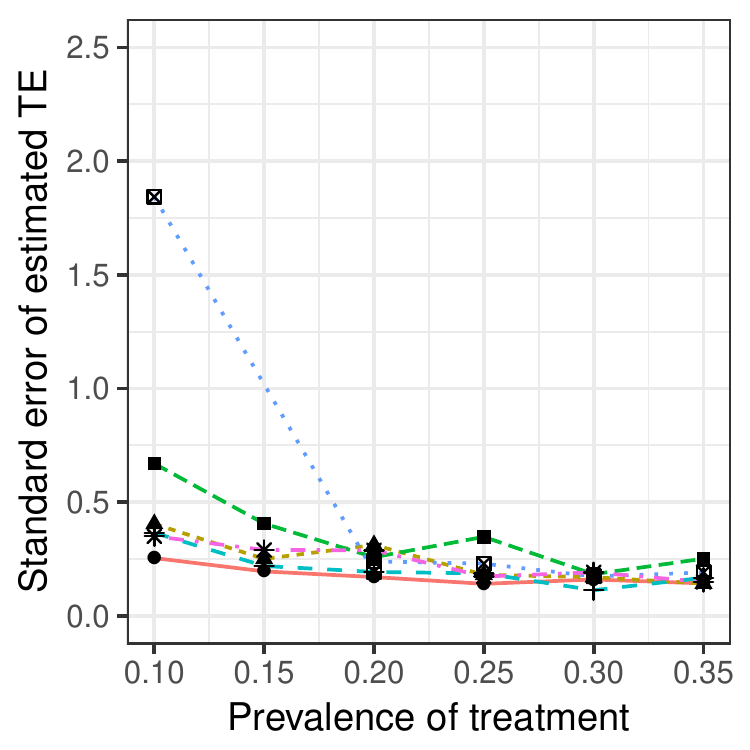}
        \caption{}
        \label{fig:std_multi}
    \end{subfigure}
    \begin{subfigure}[b]{0.25\textwidth}
        \includegraphics[height = 3.5cm, width = 3.5cm]{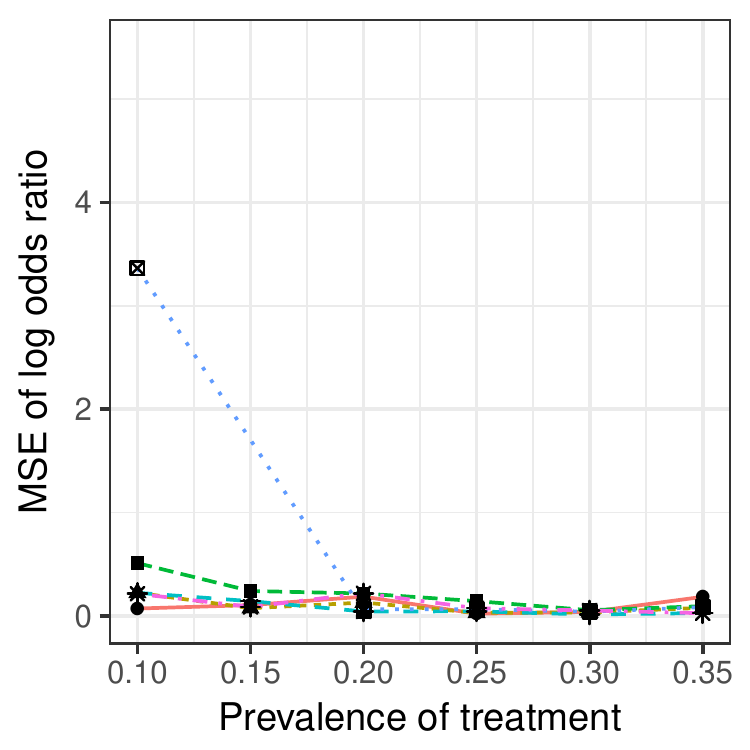}
        \caption{}
        \label{fig:mse_multi}
    \end{subfigure}
        \begin{subfigure}[b]{0.1\textwidth}
        \includegraphics[height = 2.7cm, width = 1.2cm]{name.pdf} \vspace{1cm}
    \end{subfigure}
\end{center}    
\caption{Treatment effect: log odds ratio, standard deviation of estimated log odds ratio and mean squared error of log odds ratio under multivariate normally distributed covariates.}  \label{fig:multi}
\end{figure}

Figure \ref{fig:multi} presents log odds ratio, standard deviation and mean square error of log odds ratio when the pretreatment covariates were multivariate normally distributed. The estimated treatment effect is reported in Figure \ref{fig:odds_multi}. We see that nearest neighbor matching with replacement performs better than nearest neighbor matching without replacement. Largest caliper matching performed well through different treatment prevalence. The standard deviation and mean square error of the estimated log odds ratio are reported in Figure \ref{fig:std_multi} and \ref{fig:mse_multi}, respectively. Optimal matching and full matching showed less standard deviation and less mean square error in this case. The standard deviation was high for nearest neighbor matching with replacement when the treatment prevalence is low. Genetic matching performed better than any other methods when covariates were multivariate normally distributed. 

\begin{figure}[!ht]
\begin{center}
    \begin{subfigure}[b]{0.25\textwidth}
        \includegraphics[height = 3.5cm, width = 3.5cm]{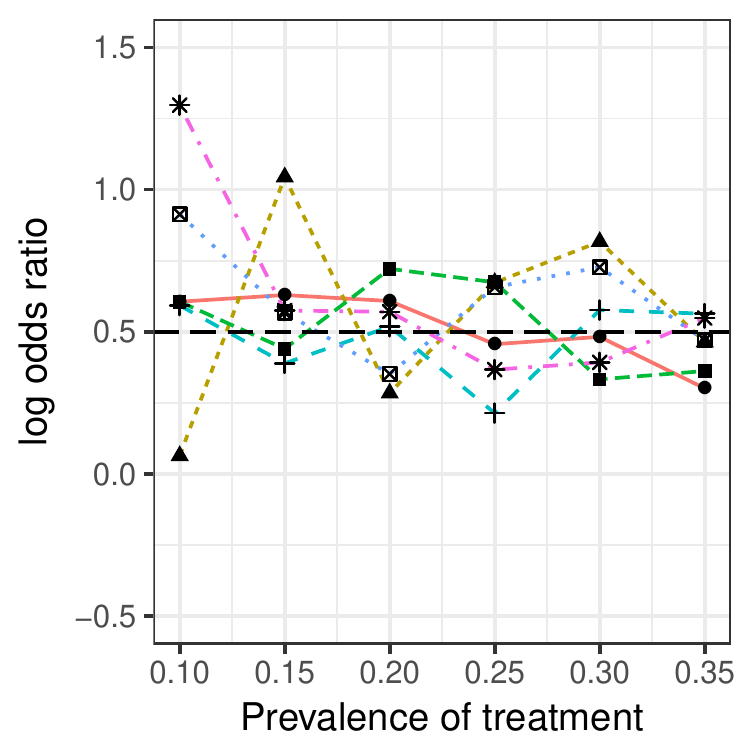}
        \caption{}
        \label{fig:odds_mulbi}
    \end{subfigure}
    ~ 
    \begin{subfigure}[b]{0.25\textwidth}
        \includegraphics[height = 3.5cm, width = 3.5cm]{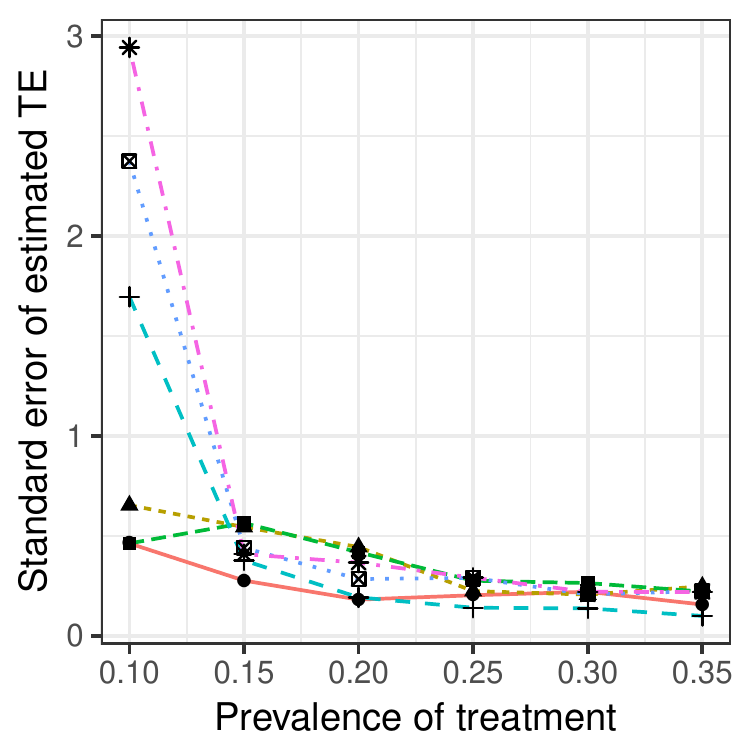}
        \caption{}
        \label{fig:std_mulbi}
    \end{subfigure}
    \begin{subfigure}[b]{0.25\textwidth}
        \includegraphics[height = 3.5cm, width = 3.5cm]{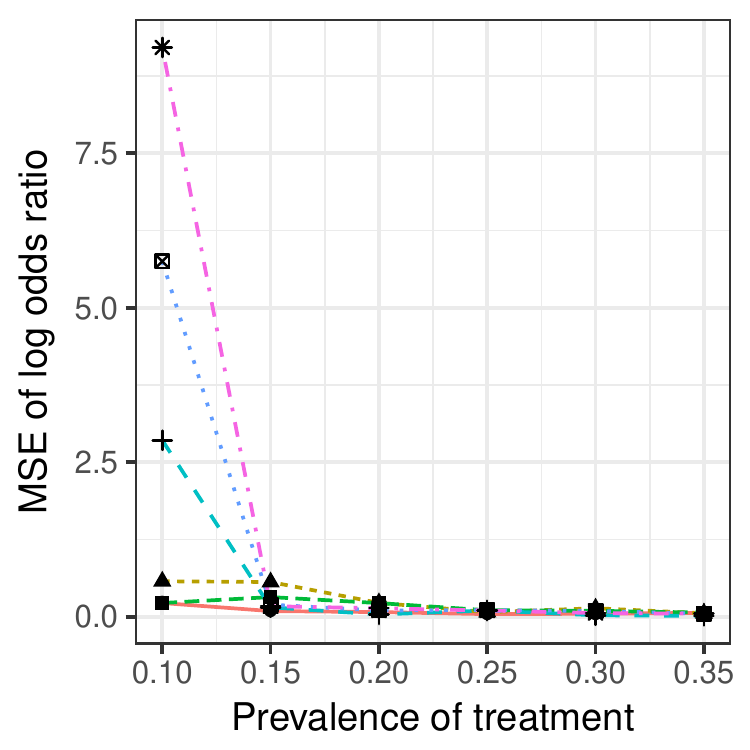}
        \caption{}
        \label{fig:mse_mulbi}
    \end{subfigure}
        \begin{subfigure}[b]{0.1\textwidth}
        \includegraphics[height = 2.7cm, width = 1.2cm]{name.pdf} \vspace{1cm}
    \end{subfigure}
\end{center}    
\caption{Treatment effect: log odds ratio, standard deviation of estimated log odds ratio and mean squared error of log odds ratio under both normally distributed and binary distributed covariates.}  \label{fig:mulbi}
\end{figure}

In Figure \ref{fig:mulbi} we report the log odds ratio, standard deviation and mean square error of the log odds ratio when the pretreatment covariates were both normally and binary distributed. Figure \ref{fig:odds_mulbi} shows the bias of the methods under different treatment prevalence. Largest caliper matching performed  consistent over different prevalence of treatment. Figure \ref{fig:std_mulbi} and \ref{fig:mse_mulbi} show the standard deviation and mean square error of the log odds ratio, respectively. Optimal matching and nearest neighbor matching with replacement had low precision in presence of low treatment prevalence. Both 1:3 full matching with calipers and largest caliper matching performed better than other matching methods.

\begin{figure}[!ht]
\begin{center}
    \begin{subfigure}[b]{0.25\textwidth}
        \includegraphics[height = 3.5cm, width = 3.5cm]{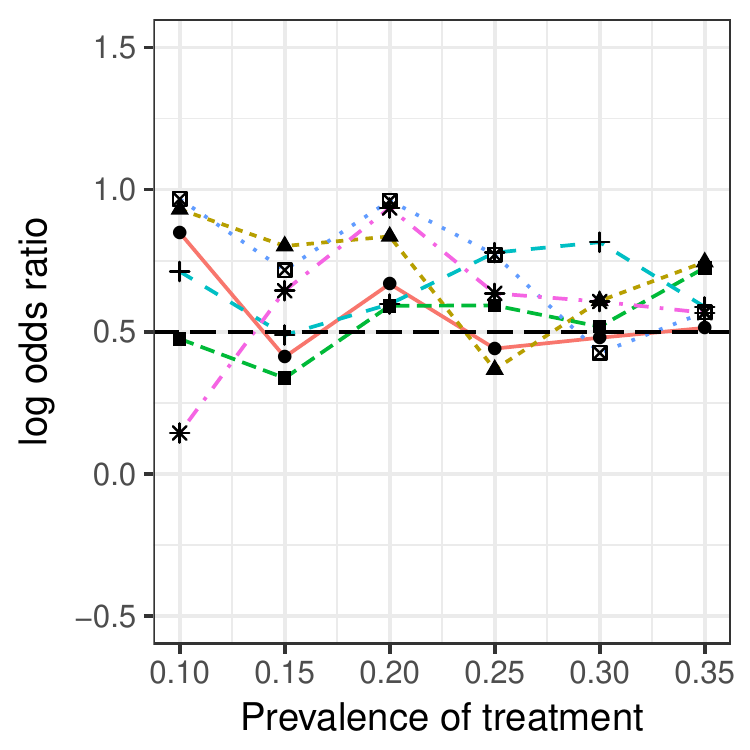}
        \caption{}
        \label{fig:odds_binary}
    \end{subfigure}
    ~ 
    \begin{subfigure}[b]{0.25\textwidth}
        \includegraphics[height = 3.5cm, width = 3.5cm]{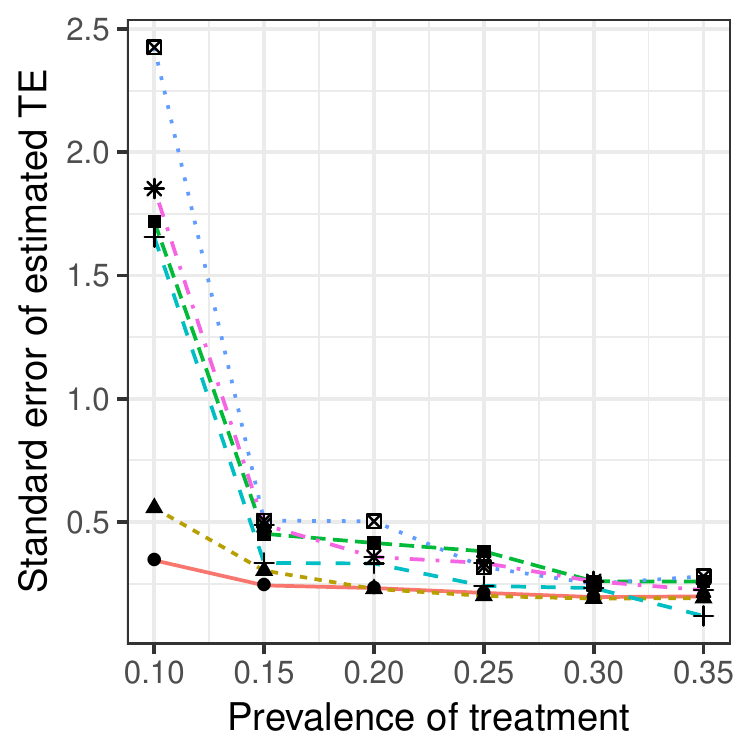}
        \caption{}
        \label{fig:std_binary}
    \end{subfigure}
    \begin{subfigure}[b]{0.25\textwidth}
        \includegraphics[height = 3.5cm, width = 3.5cm]{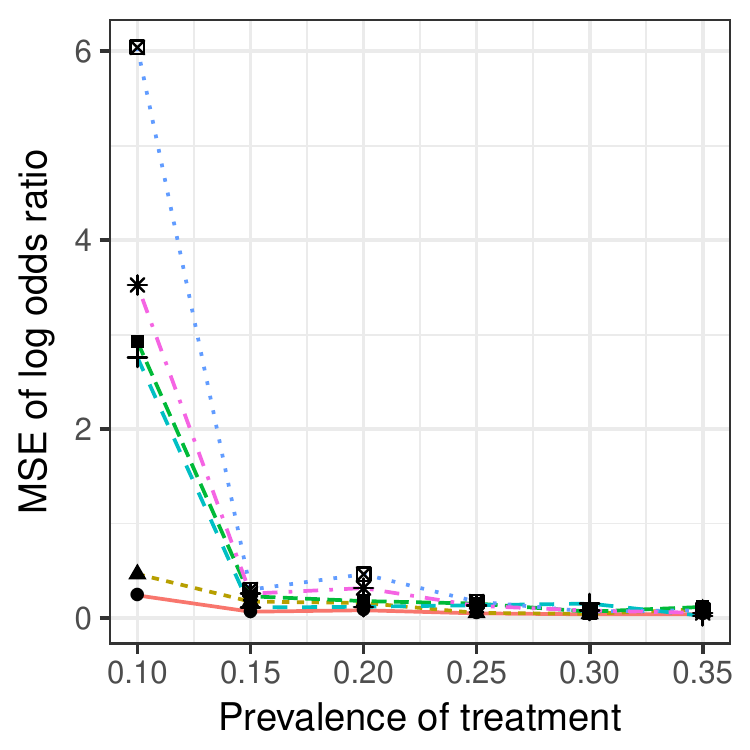}
        \caption{}
        \label{fig:mse_binary}
    \end{subfigure}
        \begin{subfigure}[b]{0.1\textwidth}
        \includegraphics[height = 2.7cm, width = 1.2cm]{name.pdf} \vspace{1cm}
    \end{subfigure}
\end{center}    
\caption{Treatment effect: log odds ratio, standard deviation of estimated log odds ratio and mean squared error of log odds ratio under binary distributed covariates.}  \label{fig:binary}
\end{figure}

In Figure \ref{fig:binary} we report the log odds ratio, standard deviation and mean square error of the log odds ratio when pretreatment covariates were independently binary distributed. Figure \ref{fig:odds_binary} shows the bias of the methods under different treatment prevalence. Both genetic matching and 1:3 full matching performed better than other methods in presence of low treatment prevalence. Figure \ref{fig:std_binary} and \ref{fig:mse_binary} show the standard deviation and mean square error of the estimated log odds ratio, respectively. Nearest neighbor with replacement performed worse in this case.

\section{Case Study} \label{casestudy}
The study analyzed Right Heart Catheterization (RHC) study to investigate whether RHC led to increase odds of severe clinical outcomes, previously analyzed by several authors \citep{Connors1996effectiveness, Imbens2001estimation}.  %
Also, it applied six matching methods on the effectiveness of RHC using data from the Study to Understand Prognoses and Preferences for Outcomes and Risks of Treatments (SUPPORT). The RHC study collected on hospitalized adult patients at 5 medical centers in the U.S. Based on information from a panel of experts a rich set of variables relating to the decision to perform the RHC and outcome. \citet{Connors1996effectiveness} found that after adjusting for ignorable treatment assignment conditional on a range of covariates, RHC appeared to lead to increase clinical death. This conclusion contradicted popular perception that RHC patients had less risk of clinical outcome. A detailed description of the study can be found in \citet{Connors1996effectiveness} and \citet{Imbens2001estimation}.

The data had 5735 subjects, 2184 treated patients and 3551 control patients. For each subject, treatment status was observed equal to 1 if RHC was applied within 24 hours of admission, and 0 otherwise. Clinical outcome was an indicator for death within 30 days. There were 68\% of the RHC patients that had clinical outcome compared to 63\% of the No RHC patients. Fifty covariates were considered for covariate matching based on the covariates that are associated with the both RHC and clinical outcome.

\begin{figure}
\centering
\includegraphics[height=16cm, width=15cm]{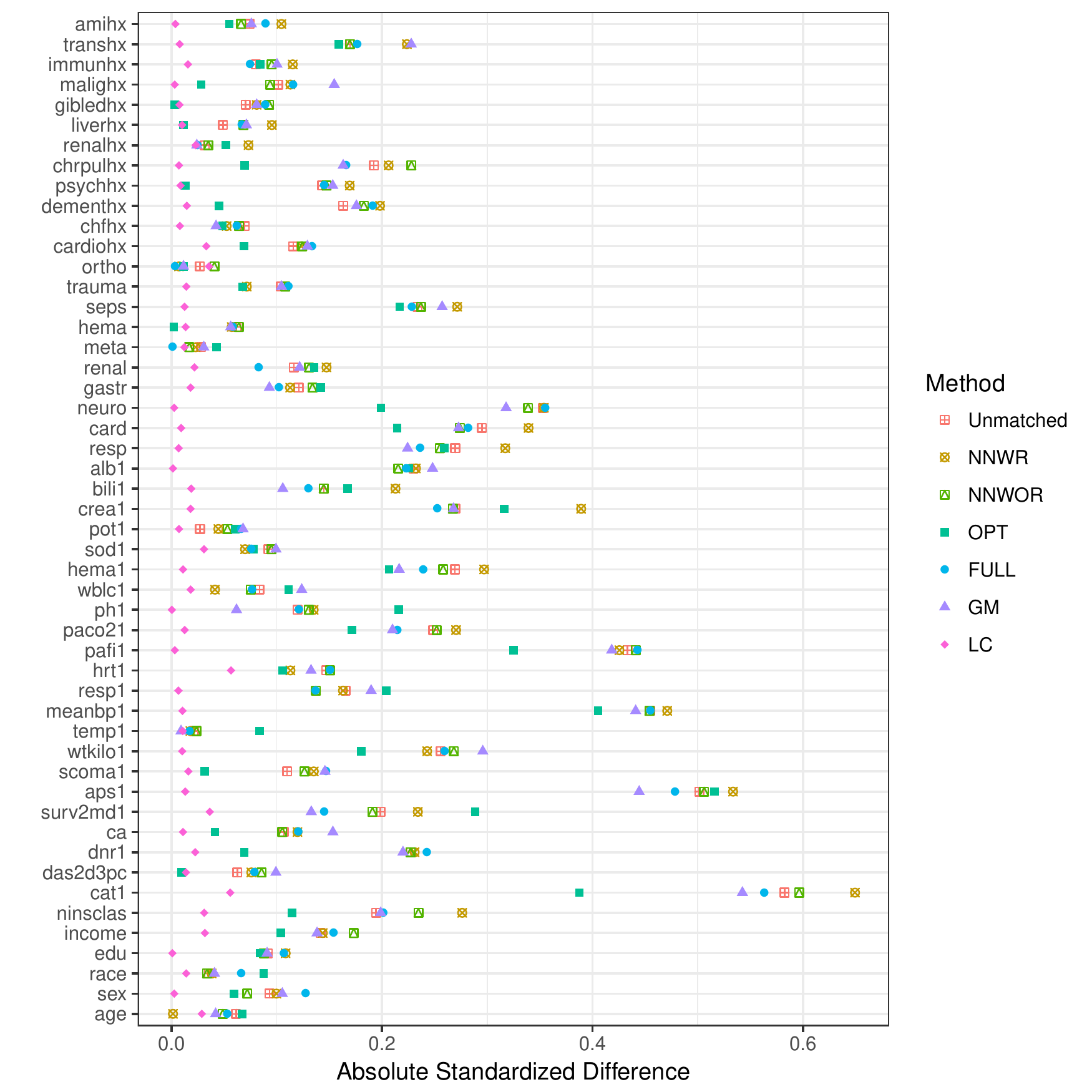}
\caption{Covariate imbalance between treated/control subjects. The dotplot (a Love plot) shows the absolute standardized differences for unmatched and six matched samples.} \label{imbalance}
\end{figure}

In unmatched data, out of 50 covariates there were 32 covariates that had absolute standardized differences were more than $0.1$. NNWR and NNWOR had 34 and 31 covariates that had absolute standardized differences more than $0.1$. OPT performed better than nearest neighbor matching in terms of reducing covariate imbalance. LC successfully reduced all the covariate imbalances in the data and the result were consistent with other matching methods. Figure \ref{imbalance} reports the standardized difference for each of the 50 covariates in the matched and unmatched data. 

We analyzed the unmatched data and the matched samples obtained from six matching methods. Table \ref{tabout} shows the outcome analysis of the SUPPORT data. The second column presents the odds ratios of the analyses. We report that RHC was significant at 5\% level of significance under all matching methods. 
\begin{table}[!ht]
\centering
\begin{tabular}{l c c c}
\hline
Method & OR & 2.5\% & 97.5\% \\ 
\hline
Unmatched & 1.252& 1.119&  1.402 \\
NNWR & 1.267 & 1.074& 1.492\\
NNWOR & 1.215& 1.068&  1.383 \\
OPT &  1.444 & 1.364 &  1.747\\
FULL & 1.167& 1.023&  1.333\\
GM & 1.243& 1.097&  1.409 \\
LC & 1.276& 1.121&  1.452 \\
\hline
\end{tabular}
\caption{Odds ratio of RHC group compare to No RHC group with 95\% confidence interval.}\label{tabout}
\end{table}%


\section{Conclusion} \label{conclusion}
The article discusses a new matching technique and compare the relative performance of the method with current existing methods under different Monte Carlo simulations setup. In this section, we briefly discuss the results.

In general, we observed several important facts that researchers need to consider in employing these matching methods. 
First, as the prevalence of the treated subjects increased from 10\% to 35\% in data, all methods tend to estimate unbiased estimate in the data and both standard deviation and mean square error of the estimates started to decrease. 
Second, full matching (in our case 1:3 with caliper) imposed more subjects than other methods---tended to result more precise estimates compared with the other matching methods. Note that full matching would perform better to reduce the covariate bias in the outcome analysis but could worsen covariate imbalance. 
Third, the choice between nearest neighbor with replacement and nearest neighbor matching without replacement reflected a bias-variance trade-off. In general, the nearest neighbor with replacement had lowest bias but higher variance compares to nearest neighbor without replacement. Some authors demonstrated this fact---matching with replacement produces matches of higher quality than matching without replacement by increasing the set of possible matches but have greater variability \citep{Abadie2006}. 
Fourth, when covariates have multivariate normally distributed covariates---genetic matching tended to have a performance that was at least as good as any of the competing methods. 
Fifth, we used Mahalanobis distance metric for nearest neighbor with replacement, nearest neighbor without replacement, optimal matching and full matching. In simulation we observed that for small number of covariates (in our case we considered five covariates) Mahalanobis distance metric performs much better than propensity score matching. 
Sixth, largest caliper matching considers an amount of covariate balance first then maximize the number of units within that balance whereas other methods iterate to improve the covariate balance. Though one can consider an optimal imbalance for largest caliper, it is recommended to use prespecified balance  on important covariates only. 
Finally, our conclusions might be restricted to our simulation scenarios and might not apply to situations not represented by our simulated data.

The quantity of interest always depends on researcher objectives---that need to setup before analysis. If the number of control subjects are insufficient then nearest neighbor without replacement can result in exclusion of some treated subjects from the matched sample. \citet{Rosenbaum1985Cons} used the term `bias due to incomplete matching' to describe the bias that arises when treated subjects are excluded from the matched sample. 
In many real application, it is could be beneficial to discard some treated subjects without good match to obtain a good covariate balance.
If a matching method discards treated subjects---the quantity of interest is no longer ATT. Since, in our simulation we considered the treatment prevalence maximum of 35\%, our quantity of interest for all matching methods was ATT. 

The results show that largest caliper matching performed fair under different setup. In presence of large number covariates, we recommend to use all the covariates that are important for both treatment assignment and outcome. Unnecessary inclusion of covariates in the matching methods could reduce the performance of the methods \citep{Stuart2010Matching}. 
Besides employing caliper on covariates---adding calipers of width equal to 0.2 of the standard deviation of the logit of the propensity score for largest caliper matching in large data could make better performance.  In this article, the analyses was conducted as a post-stratified sample---all the formed clusters were given weight to estimate ATT. 
In methodological literature, researchers have conducted substantial research on methods to estimate treatment effects. Besides, computationally they are very convenient---there are several {\tt R} packages available for matching methods, e.g. {\tt Matching}, {\tt MatchIt} and {\tt optmatch}.

We like to note certain attentions for the users of largest caliper matching. 
First, in largest caliper matching the analysis is sensitive to the choice of the caliper that could make substantial difference in matched sample. One choice of the caliper could be, consider only the important covariates that have higher standardized difference than a tolerance level.  
Second, a tighter caliper leads to reduce bias and make good matches but could discard those treated subjects that do not have good matches. 
Third, largest caliper matching ensures that there is at least one match for all treated subjects when the quantity of interest is ATT. Fourth, largest caliper matching is fast for a given amount of imbalance that researchers want to accept for a covariate.  For SUPPORT data set, our largest caliper matching took 2.7 seconds to to run on a desktop computer with 2.7 GHz Intel Core i7 processor and 16.0 GB RAM. Finally, largest caliper matching forms a good match sample that forms a cluster of homogeneous subjects. It successfully discards the control subjects that could increase the imbalance in the data. Combining these characteristics, largest caliper matching is a very computationally efficient and convenient matching method.

\subsubsection*{Acknowledgements:}
The author would like to thank all the fellows who reviewed the article from Kansas State University and The University of Iowa.



\end{document}